\documentclass[10pt,conference]{IEEEtran}
\usepackage[utf8]{inputenc}

\IEEEoverridecommandlockouts

\usepackage{pgfplots}
\usepackage{pgf-pie}
\usepackage{graphicx}
\usepackage{fancyvrb}
\newcommand{\code}[1]{\texttt{#1}}

\newcommand{\hide}[1]{}
\newcommand{\anon}[2]{#1}

\begin{document}

\newcommand{\arith}{\textsc{ArithSum}}
\newcommand{\var}{\textsc{Variance}}
\newcommand{\prim}{\textsc{Prime}}
\newcommand{\sort}{\textsc{Sort}}
\newcommand{\rtrig}{\textsc{RightTrig}}
\newcommand{\parln}{\textsc{Parallel}}

\title{Does Code Structure Affect Comprehension?\\
On Using and Naming Intermediate Variables
\thanks{\anon{This research was supported by the ISRAEL SCIENCE FOUNDATION (grant no.\ 832/18).}{}}
}

\author{\anon{\IEEEauthorblockN{Roee Cates$^*$\thanks{$^*$Authors contributed equally.} ~~~~~ Nadav Yunik$^*$ ~~~~~ Dror G. Feitelson}
\IEEEauthorblockA{Department of Computer Science\\
The Hebrew University of Jerusalem, 91904 Jerusalem, Israel}}
{Authors Anonymized}}

\maketitle

\begin{abstract}
Intermediate variables can be used to break complex expressions into more manageable smaller expressions, which may be easier to understand.
But it is unclear when and whether this actually helps.
We conducted an experiment in which subjects read 6 mathematical functions and were supposed to give them meaningful names.
113 subjects participated, of which 58\% had 3 or more years of programming work experience.
Each function had 3 versions: using a compound expression, using intermediate variables with meaningless names, or using intermediate variables with meaningful names.
The results were that in only one case there was a significant difference between the two extreme versions, in favor of the one with intermediate variables with meaningful names.
This case was the function that was the hardest to understand to begin with.
In two additional cases using intermediate variables with meaningless names appears to have caused a slight decrease in understanding.
In all other cases the code structure did not make much of a difference.
As it is hard to anticipate what others will find difficult to understand, the conclusion is that using intermediate variables is generally desirable.
However, this recommendation hinges on giving them good names.
\end{abstract}


\section{Introduction}

The ability to comprehend other software engineers' code is a key ingredient of software maintenance.
Whether it is debugging and fixing code, adapting and modifying it, reusing code or leveraging its functionality, comprehension is an imperative stage in the process.
Both the comprehension itself and the time it takes to understand some functionality can be hampered by the code’s complexity and lack of readability.

In many cases, the same functionality can be expressed in different ways.
And some structures, more than others, may increase engineers’ cognitive load.
One example is the use of compound expressions as opposed to breaking the expression into separate sub-expressions, and tying them together using intermediate variables.
As a simple example, the distance between two points \code{A} and \code{B} can be calculated as
\begin{Verbatim}[frame=single]
d = sqrt( (A.x-B.x)**2 + (A.y-B.y)**2 )
\end{Verbatim}
Or alternatively as
\begin{Verbatim}[frame=single]
dx = A.x - B.x
dy = A.y - B.y
d = sqrt( dx**2 + dy**2 )
\end{Verbatim}
The first version contains a larger compound expression, but overall it is shorter.
In the second version each expression is simpler, and has a name which reflects its essence. But there are three of them, and one needs to mentally connect them to get the full picture.
The question, then, is whether using simpler expressions connected by intermediate variables is advantageous, or alternatively whether sticking with the shorter more condensed version is perhaps better.

The idea of using intermediate (temporary) variables like this is not new.
In fact, the change from the first to the second version above is an instance of an ``extract variable'' refactoring (also known as ``introduce explaining variable'') \cite{fowler:refactor}.
At the same time, changing from the second version to the first is an instance of an ``inline variable'' refactoring.
The fact that both refactorings exist implies that both are thought to be beneficial in certain conditions.
Our goal is to study this experimentally, in the context of code comprehension.

In our experiments the participants are required to understand short functions which are presented in either of three styles:
using compound expressions,
with variables extracted but given meaningless names,
and with variables extracted and given meaningful names.
This design enables us to see whether effects are due to the change in structure or to the introduction of the names.

The results of executing such an experiment based on 6 mathematical functions suggest that the meaningful names are the more important factor.
Introducing intermediate variables with meaningless names might even lead to making more comprehension errors.
But we also find that intermediate variables make a difference only in relatively complex cases, where comprehending the code is a challenge.
If it is straightforward enough, it hardly matters whether intermediate variables are used or not.
Since it is hard to anticipate accurately what others will find hard, using intermediate variables with meaningful names seems the safe approach.

\section{Related Work}

There has been very little work on the comprehensibility of expressions, and more generally on the comprehensibility of coding structures.
Ajami et al.\ compared multiple code snippets which expressed the same functionality in different ways \cite{ajami19}.
Among other things they compared compound logical expressions with nested structures of \code{if}s.
The results were that no significant differences in comprehensibility were found.
Such structures were also considered by Wiese et al.\ \cite{wiese19}, where it was accepted that expert opinion is that compound expressions are preferable to nested \code{if} statements.
But when novices were confronted with such code differences, it did not have a significant effect on their performance, even though they said they found the compound version more readable.

Removing intermediate variables and favoring compound expressions is one of the transformations advocated in the context of ``Spartan'' coding \cite{gil17b,gil17c}.
The goal there is to reduce all code metrics, including its size (as measured by the number of characters in the program), the number of control statements, and the number of variables.
However, we do not know of any empirical studies concerning the effect of these transformations on comprehension.

There has been significantly more work on the effect of variable names.
Considerable attention has been devoted to the issue of name lengths, and in particular to whether abbreviations cause a disadvantage relative to using full words \cite{binkley09c,hofmeister19,holzmann16,lawrie06,scanniello13,schankin18}.
Another favorite topic is controlling the vocabulary used in variable names, and how this reduces ambiguity \cite{binkley15,caprile00,deissenboeck05,fakhoury18}

Feitelson et al.\ investigated how variables names are selected \cite{feitelson:names}.
One of their results was that naming is very variable, with little chance that different developers would select exactly the same name for the same variable.
Arnaoudova et al.\ investigated renaming of variables, and showed that renamings most often modify the meaning of the names \cite{arnaoudova14}.
A possible explanation for this is provided by Avidan and Feitelson, who show that names may be misleading to the point of being worse that meaningless names like consecutive letters of the alphabet \cite{avidan17}.
All these works testify to the difficulty of finding good names that will convey the desired meaning without any ambiguity.

Introducing intermediate variables with meaningful names can serve as a form of inline documentation.
Zabardast et al.\ claim that this is one of the refactorings that are effective in reducing technical debt, based on an analysis of 2286 commits in an unnamed financial services application \cite{zabardast20}.
Counsell et al.\ claim that an association exists between classes in which this refactoring was applied and defect proneness, based on 5 releases of 2 Eclipse projects \cite{counsell15}.
This may imply that the refactoring was applied to classes that were known to be problematic.
Our emphasis is on the effect on code comprehension.
In addition, we attempt to distinguish between the contribution of the restructuring of the code and the addition of a meaningful variable name.

\section{Research Questions}

Our overarching goal is to better understand how coding practices affect code comprehension.
In this paper we specifically focus on the format of mathematical functions, and in particular on the possible effect of using intermediate variables to store the results of sub-expressions.
To understand this issue we need to distinguish between two interrelated factors.
The first is the partitioning of large compound expressions into a set of smaller and simpler sub-expressions.
The second is giving meaningful names to these sub-expressions.
These considerations lead to the following two concrete research questions:
\begin{enumerate}
    \item Does compound code impair comprehension compared to partitioned code which uses intermediate variables to store sub-expression results?
    \item Given such partitioned code, do meaningful intermediate variable names support comprehension?
\end{enumerate}

\section{Methodology}

To study the effect of intermediate variables we conducted a controlled experiment.
We used 6 functions, each with 3 code versions that differed in their use of intermediate variables.
The experimental subjects were tasked with understanding the functions and giving them names.

\subsection{Selection of Functions}

We originally defined 10 fundamental mathematical functions.
The reason for choosing such functions was the desire to avoid domain knowledge issues.
Basic mathematical functions are expected to be understandable by any competent programmer.
To further reduce problems of domain knowledge, we chose functions from different fields of mathematics, such as geometry and probability.
Note too that understanding mathematical functions is highly relevant to developers, who may face the need to understand such code in the context of machine learning, data science, or statistics.

Another consideration was to include functions with different levels of  complexity.
This is somewhat ill-defined, as a function that one experimental subject finds easy to comprehend may cause another subject to struggle.
We therefore conducted a small pilot study with 3 subjects (two Computer Science BSc graduates and an algorithms developer).
This study showed that the experiment took more time than we planned, and that some of the questions were harder than we anticipated.
We therefore eliminated 4 of the functions:
a function calculating the solution to a quadratic equation, which was considered too easy, due to the iconic formula for the discriminant (b**2 - 4*a*c);
a function calculating the sum of a geometric series which was too similar to another calculating the arithmetic sum which we retained;
a function calculating the Manhattan distance between two points, that led to excessive ambiguity in the answers;
and a function calculating the cosine similarity between two vectors, which was too complex and was not understood by any of the pilot participants.

\begin{figure*}
Version 1: single compound expression
\begin{Verbatim}[frame=single]
def foo(arr):
    return sum((x - (sum(arr) / len(arr)))**2 for x in arr) / len(arr)
\end{Verbatim}

Version 2: break using intermediate variables
\begin{Verbatim}[frame=single]
def foo(arr):
    tmp1 = len(arr)
    tmp2 = sum(arr) / tmp1
    return sum((x - tmp2)**2 for x in arr) / tmp1
\end{Verbatim}

Version 3: give them meaningful names
\begin{Verbatim}[frame=single]
def foo(arr):
    n = len(arr)
    mean = sum(arr) / n
    return sum((x - mean)**2 for x in arr) / n
\end{Verbatim}

    \caption{Versions of the \var\ function}
    \label{fig:versions}
\end{figure*}

The final experiment used the remaining 6 basic mathematical functions. 
These were:
\begin{itemize}
    \item \arith: Assert that the elements of an array form an arithmetic series (i.e.\ that they are equidistant), and compute the sum of this series using the conventional formula.
    \item \var: Compute the variance of the elements of an array.
    \item \prim: Establish whether a number is a prime, by checking whether any smaller number divides it.
    \item \sort: Sort an array of numbers using bubble sort.
    \item \rtrig: Check whether a triangle has a right angle by checking whether any permutation of its sides is a Pythagorean triple.
    The check for a Pythagorean triple is done by a separate subroutine.
    \item \parln: Check whether two lines, defined by 2 points each, are parallel to each other, by checking whether they are vertical, and if not, by comparing their slopes.
\end{itemize}

\subsection{Code Versions}

For each function we created three code versions:
\begin{enumerate}
    \item The base version used a compound expression which had no intermediate variables at all.
    \item A version in which the compound expression is broken into sub-expressions, whose values were stored in intermediate variables.
    These variables were given meaningless names: \code{tmp1}, \code{tmp2}, and so on.
    \item The last version was the same as the previous version, but this time the intermediate variables were given meaningful names such as \code{mean} or \code{isGreater}.
\end{enumerate}

An example is shown in Fig.\ \ref{fig:versions}.
These are the three versions of the \var\ function (which calculates the variance of an array of numbers).
Version 1, with no intermediate variables, employs a single compound expression embedded in the \code{return} statement.
In version 2 intermediate variables are used to store the input array's length and the average of the input elements.
These variables are then used in the expression calculating the variance, thereby simplifying it.
However, their names do not reflect their meanings.
Version 3 is the same, but this time the names (\code{n} and \code{mean}) do reflect the meanings of these variables.

By using these three versions we divide the difference between using a compound expression and extracting explanatory intermediate variables into two independent parts, as required by our research questions.
The difference between versions 1 and 2 reflects a pure structural difference.
Comparing the performance on these two versions will show whether using a single compound expression is better or worse than using multiple simpler expressions.
The difference between versions 2 and 3 is a pure semantic difference, at the level perceived by a human reader, reflected in the intermediate variable names.
Comparing the performance on these two versions will show whether any changes in performance are a result of the structural change or a result of giving the intermediate variables meaningful names.

In realistic settings, we would expect functions to be similar to versions 1 or 3.
In this sense version 2 can be seen as an experimental control.
Comparing version 1 to 3 shows whether explanatory variable extraction is beneficial, and comparing both to version 2 shows whether this is due to the extraction or to the explanatory power of their names.

Note that the functions are always named \code{foo}.
This is done so as not to give away what the function does, thereby requiring the experimental subjects to read the code and try to understand it.

\subsection{The Experimental Task}

The task which subjects are required to perform is to read the functions and understand them.
Recall that all functions are presented with the name \code{foo}.
In order to demonstrate their understanding, subjects were required to suggest a better name for each function.
This task is straightforward, and it reflects understanding because one can name a function correctly if and only if it was comprehended (up to guesses).
Subjects may also answer ``?'' which indicates ``I don’t know'' to skip a question.

The problem with this task is that we are required to grade the selected names.
The grading procedure we used is described below.
As an alternative we also considered using a multiple choice question in which subjects would be required to select the best name out of a number of options.
This has the advantage of avoiding the need to judge the quality of the answers.
However, it also has the prominent drawback of guiding the subjects towards a limited set of possible alternatives.
This can suggest what to look for in the code, and lead to a strong bias in the results, which would not reflect the subjects' independent understanding of the code.

\subsection{Experiment Execution}

The experiment was conducted online, using the LimeSurvey platform.
This platform supported all the features we needed: A/B/C testing via version randomization, questions order randomization, and measuring the time it took participants to answer each question.
No time limit was imposed.

Each configuration of a function and a version forms a trial.
The subjects received the six functions one after the other and were asked to understand and name them.
The function order was randomized in order to avoid any systematic bias due to learning or fatigue.

Each subject saw only one version of the code for each function.
The version shown was chosen at random.
Thus the comparison of the results obtained for the different versions of the same function is a between-subjects comparison.
At the same time, each subject most probably saw instances of all 3 versions, thereby reducing the effect of inter-personal differences in performance.

The independent variable in this study is the code version.
In each trial we observe two dependent variables:
\begin{enumerate}
    \item The name that the subject gave the function.
    This enables us to judge whether the subject solved the task correctly and understood the function.
    \item The time it took the subject to solve the task.
    In the analysis we focus on the times of those subjects who named the functions correctly.
\end{enumerate}
Both dependent variables are expected to correlate with the difficulty of understanding a function \cite{rajlich97}.

\section{Results}

\subsection{Participants}

The experiment was conducted in two waves (initially we did not have enough subjects, so we conducted a second wave to improve statistical validity).
Participants were recruited by posting notices in various facebook groups, such as the group of Computer Science students as Hebrew University, by posting on university whatsapp groups for 3rd year and advanced students, and by personal contacts of the authors at work.
Participants were not paid, and could skip questions and leave the experiment at any time.
In total 191 subjects entered the experiment.
36 were excluded from the analysis as they did not claim experience with python, the language we used to present the code.
Of the reminder, 113 answered at least one question, and 93 completed all 6.

\begin{figure}\centering
\begin{tikzpicture}
\pie[radius=1.2, color={black!30,orange!50,orange!70,orange!90,red!60,green!50,green},
    sum=auto, text=pin]
    {3/\small\textsf{none},
     8/\small\textsf{BSc year 1},
     12/\small\textsf{BSc year 2},
     29/\small\textsf{BSc year 3},
     27/\small\textsf{have BSc},
     24/\small\textsf{MSc student},
     10/\small\textsf{have MSc}}
\end{tikzpicture}
    \caption{Academic beckground of survey participants.}
    \label{fig:studies}
\end{figure}
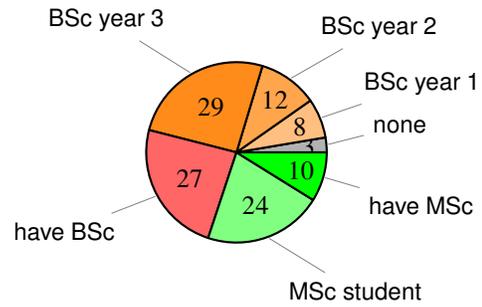

\begin{figure}\centering
\begin{tikzpicture}
\pie[radius=1.2, color={black!30,orange!70,cyan!50,blue!70},
     sum=auto, text=pin]
    {15/\small\textsf{none},
     33/\small\textsf{1-2},
     37/\small\textsf{3-5},
     28/\small\textsf{$\ge$6}}
\end{tikzpicture}
    \caption{Years of programming work experience of survey participants.}
    \label{fig:vetek}
\end{figure}
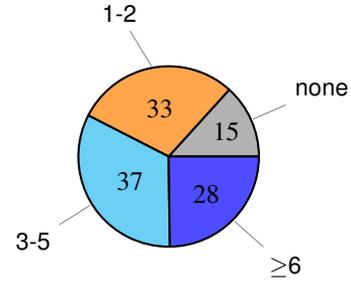

The academic background of the participants is shown in Fig.\ \ref{fig:studies}.
Of those who answered, less than half did not have a first degree:
2.7\% had no academic background, and 43.4\% were BSc students (or which the majority, 25.7\%, were in their third and final year).
23.9\% had completed their BSc,
another 21.2\% were MSc students, and 8.8\% had completed their MSc.

The programming work experience of the participants is shown in Fig.\ \ref{fig:vetek}.
13.3\% had none,
29.2\% had up to 2 years of experience,
and 57.5\% 3 years of experience or more.
Together these characteristics indicate that, while most of our experimental subjects have an academic background, only a minority are inexperienced students.
And they can be expected to be capable of understanding the codes we use in the experiment.

\subsection{Judging Name Correctness}

Successful naming was judged by inclusion of specific keywords (or concepts) in the given function names.
This was checked manually, thereby avoiding difficulties resulting from different naming styles (camelCase or using under\_scores), spelling errors, or using synonyms.

The procedure to assess name correctness was as follows:
\begin{enumerate}
    \item One of the authors drew up a suggested key for what constitutes a correct name in each case.
    \item All 3 authors independently went over all the names given to all the functions, tagging them as correct or not based on the key.
    A special tag was used to indicate cases that should be discussed.
    \item We discussed cases that were flagged for discussion or where we had given conflicting judgements, and reached a joint agreement.
    There were 38 such individual cases from a total of 601 names.
\end{enumerate}

The final keys used for the different functions were as follows.
Examples of names from the experiment (identified by \code{typewriter font}) are shown as written, including typos.
\begin{itemize}
\item \arith:
    We required the words ``arithmetic'' and ``sum'' or equivalent.
    Thus names such as \code{sum\_arithmetic\_progresion}, \code{sum\_linear\_ array}, and \code{SumOfListWithFixedIntervals} were accepted as correct.
    The vast majority of names given also included ``series'' or ``progression'', but we decided not to require this.
    A unique special case was \code{Get\_gaus\_series\_sum}, which was accepted based on the story of Gauss coming up with this formula as a schoolboy.
    Names such as \code{sumSeries} were not accepted due to being too general, and names like \code{is\_equal\_diff} or \code{calc\_arithmetic\_progresion} were rejected because they did not include the summing.
    \code{sumSteps} was rejected because it implies summing the differences between the elements rather than the elements themselves.
\item \var:
    We initially required the word ``variance'', or an abbreviation such as ``var''.
    After discussing this we decided to also accept names implying ``standard deviation'' (including just \code{std}) despite the fact that the code did not include a square-root operation, as they too
    gave evidence to understanding the essence of the calculation.
    However, \code{mse} (mean standard error) was rejected despite being close, as were names like \code{avrage\_sqr\_distance}.
    We also rejected \code{var\_iterator} even though the variance part was understood correctly, because this does not function as an iterator.
\item \prim:
    We initially required the name \code{isPrime} or close equivalent.
    However, after discussion, we decided to also accept names indicating a check for being divisible, as they too indicate an understanding of the essence while the error is only in the direction of the check.
    At the same time we rejected names such as the explicit \code{check\_n\_mod\_number\_smaller\_is\_zero} which lacks an indication of actual understanding.
\item \sort:
    We required ``sort'' and did not insist on identifying the specific algorithm.
    Thus wrong identifications of the algorithm, such as \code{InsertionSort}, were accepted.
\item \rtrig:
    In this case many diverse names were used, so we manually verified that they were correct.
    Examples include \code{90\_degrees\_in\_triangle} and \code{is\_right\_triangle}.
    References to ``straight angle'' were also accepted\anon{, as this is the actual translation of the Hebrew term for ``right angle''}{}.
    We also decided to accept names like \code{IsPythagorianTriplet}, even though strictly speaking this actually describes the service routine used by the function.
    As the service routine appeared first, subjects might have thought they were required to name this routine.
    Names like \code{is\_equilateral\_triangle} were rejected.
    \code{isPerp} was rejected as being too general.
\item \parln:
    We required either an indication of ``parallel'', or the combination ``same'' and ``slopes''.
    Example include \code{AreParallel} and \code{are\_of\_equal\_slopes}.
    We also accepted \code{isParallelogram} and even \code{isRectangle} despite not being strictly correct.
    We did not accept \code{isBothVertical} as being too specific.
\end{itemize}

\subsection{Success Rate}

\hide{
\begin{table}\centering
    \caption{Correctness results and p-values for comparisons between versions.}
    \label{tab:results}
\begin{tabular}{|l|ccc|ccc|}
\hline
& \multicolumn{3}{c|}{Fraction correct} & \multicolumn{3}{c|}{p-value} \\
Function    & v1    & v2    & v3    & v1-v2 & v2-v3 & v1-v3 \\
\hline
\arith      & 0.545 & 0.444 & 0.594 & 0.405 & 0.222 & 0.696 \\
\var        & 0.636 & 0.667 & 0.714 & 0.789 & 0.681 & 0.522 \\
\prim       & 0.840 & 0.884 & 0.889 & 0.611 & 0.943 & 0.581 \\
\sort       & 0.941 & 0.968 & 0.937 & 0.614 & 0.577 & 0.950 \\
\rtrig      & 0.325 & 0.481 & 0.719 & 0.201 & 0.065 & 0.001 \\
\parln      & 0.618 & 0.500 & 0.775 & 0.366 & 0.022 & 0.143 \\
\hline
All Combined   & 0.633 & 0.673 & 0.775 & 0.399 & 0.023 & 0.002 \\
\hline
\end{tabular}
\end{table}
}

\begin{figure}
    \centering
    \includegraphics[width=0.48\textwidth]{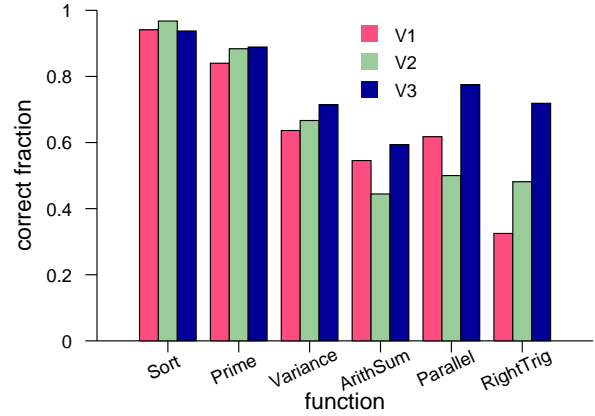}
    \caption{Fraction of correct answers for each question version.}
    \label{fig:correct}
\end{figure}

The success rate is the number of correct responses divided by the number of total responses.
The results for all versions of all functions are shown in
Fig.\ \ref{fig:correct}.
Three patterns stand out:
\begin{itemize}
    \item The first pattern is one of no large differences.
    This was the case for the \sort, \prim, and \var\ functions.

    \item In the \arith\ function and the \parln\ function the middle version (with intermediate variables with meaningless names) caused a noticeable \emph{decrease} in the success rate.
    But the third version, with the meaningful names, led to success rates that were higher than those of the original compound-expression version.

    \item Finally, in only one function (\rtrig) was there a substantial progression between the 3 versions.
    Breaking the compound expression led to a noticeable increase in the success rate, and adding meaningful names led to another noticeable increase.
\end{itemize}

We tested whether any of these observations is statistically significant using the `N-1' Chi-squared test \cite{campbell07}.
Only two of the results were found to be statistically significant at the 0.05 level.
These were the difference between versions 2 and 3 in the \parln\ question, and the difference between versions 1 and 3 in the \rtrig\ question.
Importantly, this last difference stays significant also after Bonferroni correction for multiple tests.
We conclude that \rtrig\ is the only function where inserting  explanatory variables really helped.

Note that while only one function produced a statistically significant difference, in 4 others the difference was in the same direction (more correct answers to version 3 than to version 1).
A possible alternative analysis is then to compare version 1 to version 3 using the evidence from all 6 functions together.
We did this in two separate ways.
The first was to use the `N-1' Chi-squared test as above.
The result was highly significant (p=0.002), and stays significant after Bonferroni correction.
The second is to compare the correctness results for the 6 functions using a Wilcoxon signed-rank test.
As there is only one case where the different was in favor of version 1, and this was the smallest difference (and therefore had the lowest rank), this test too indicated that there exists a statistically significant difference between versions 1 and 3.

To summarize, using intermediate variables with meaningful names indeed appears to lead to better comprehension, but this is largely due to one of the 6 functions we used.

\subsection{Time to Correct Answer}

\begin{figure*}
    \centering
    \includegraphics[width=0.9\textwidth]{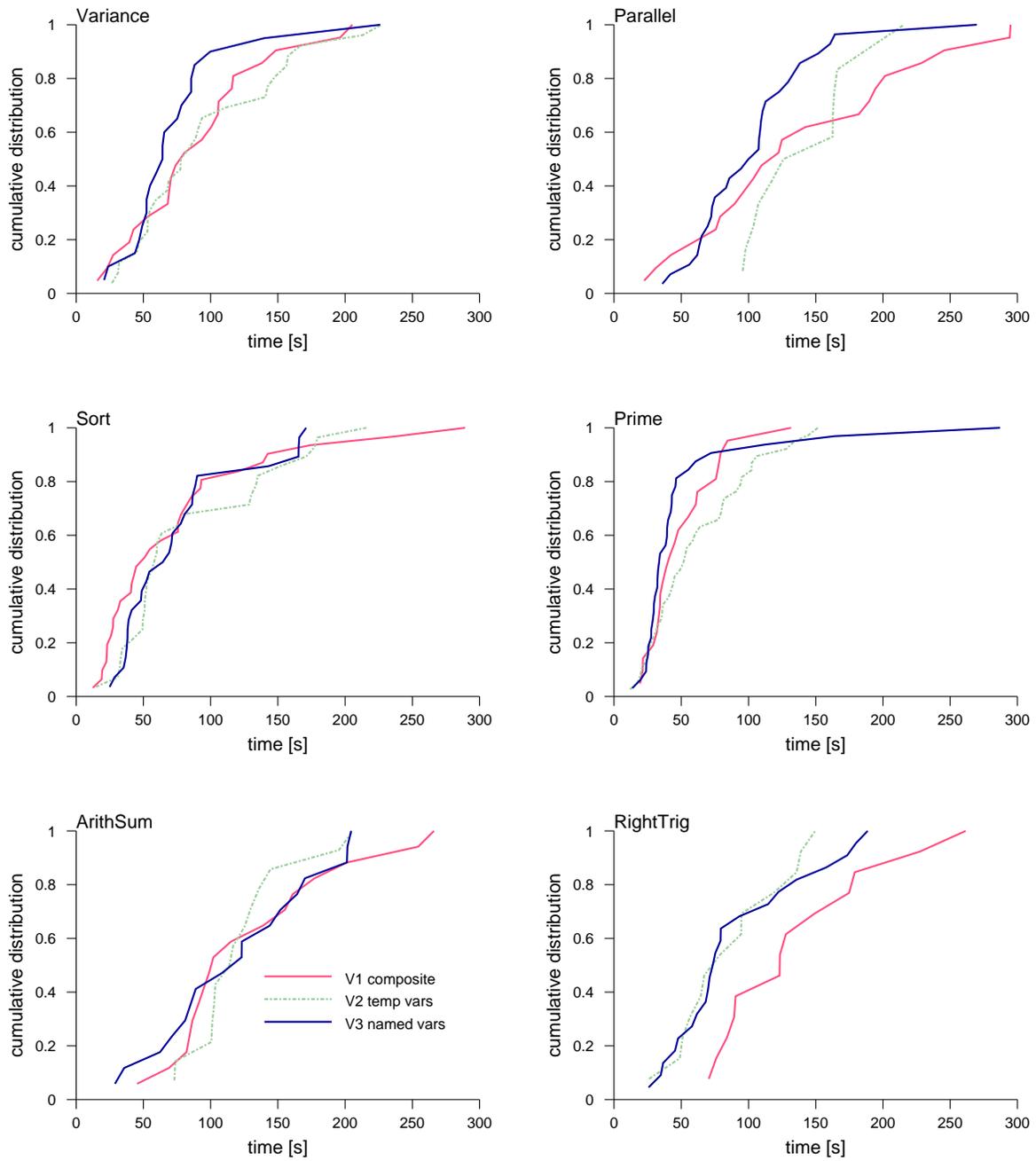}
    \caption{Results of time to correct answer for the different functions and versions.}
    \label{fig:cfds}
\end{figure*}

The second dependent variable we measured was the time to achieve a correct answer.
The resulting cumulative distribution functions (CDFs) for all correct answers to the 3 versions of each question are shown in Fig.\ \ref{fig:cfds}.
We cut the distributions at 300 seconds (5 minutes) and excluded results that took longer.
The justification is that they were obvious outliers, based on the knee-shape of the CDFs: nearly all the data was concentrated in the range of low values, and only few dispersed values were higher.
There were 15 such cases from a total of 417 correct answers.

Comparing the CDFs of versions 1 and 3, there are 3 functions for which they are very close to each other: \arith, \sort, and \prim.
For these functions it seems that adding intermediate variables with meaningful names did not lead to a systematic change in the time to achieve a correct answer.

In the other 3 functions the CDF of version 1 (compound expression) seems to dominate the CDF of version 3 (meaningful names) for at least a large part of the range.
Being a dominant distribution means that for every value the probability to observe this value or a larger one is higher than for the other distribution.
In terms of CDFs, this is manifest by the CDF being shifted to the right.
For \var\ it happens from the 30th to the 95th percentile,
for \parln\ this happens in the range from the 20th to percentile to the end, and for \rtrig\ it occurs throughout the full range.

It is also interesting to observe the relationship between the CDF of version 2 and the other two versions.
In two of the functions, \arith\ and \sort, it is similar to the CDFs of the other versions (which are also similar to each other).
In three others the CDF of the version 2 seems to dominate the other two, and is closer to the CDF of version 1: in \prim, \var, and \parln.
This means that just adding intermediate variables but without meaningful names led to spending a bit more time.
Surprisingly, in the last function, \rtrig, the CDF of version 2 is similar to that of version 3, and is dominated by that of version 1.
So in this case using intermediate variables --- even when they had meaningless names --- led to spending a little \emph{less} time.

\hide{
\begin{table}\centering
    \caption{Mann Whitney estimated shift and p-values.}
    \label{tab:results}
\begin{tabular}{|l|cc|cc|cc|}
\hline
& \multicolumn{2}{c|}{v1-v2} & \multicolumn{2}{c|}{v2-v3} & \multicolumn{2}{c|}{v1-v3} \\
Function & shift & p-val & shift & p-val & shift & p-val \\
\hline
\arith   &  -4.6 & 0.769 &  2.4 & 0.860 &  6.7 & 0.786 \\
\var     &  -5.0 & 0.791 & 15.5 & 0.142 & 16.8 & 0.218 \\
\prim    &  -7.5 & 0.310 & 12.9 & 0.034 &  6.1 & 0.195 \\
\sort    & -13.1 & 0.176 &  3.9 & 0.751 & -9.7 & 0.315 \\
\rtrig   &  40.1 & 0.029 & -0.6 & 0.987 & 44.1 & 0.012 \\
\parln   & -13.7 & 0.671 & 41.5 & 0.010 & 26.7 & 0.156 \\
\hline
\end{tabular}
\end{table}
}

We tested the differences between the distributions using the Mann-Whitney non-parametric test, as most of the distributions appear to be skewed.
4 differences were found to be statistically significant at the 0.05 level:
\prim\ versions 2 and 3,
\rtrig\ version 1 relative to versions 2 and 3,
and \parln\ versions 2 and 3.
However, none of them remain significant after Bonferroni correction.
Note, however, that according to the binomial distribution the probability of 4 or more successes out of 18, when the probability of each one is only 0.05, is 0.011.
It is therefore unlikely that all 4 results are spurious.

\subsection{Semantic Focus of Names}

While this was not the original motivation for our study, our results provide unique data about how different developers name the same function.
This can be used to analyze what concepts they choose to include in the names, and what words they use to represent these concepts.

\begin{table*}\centering
    \caption{Semantic focus results for naming the different functions in the experiment.}
    \label{tab:focus}
    \begin{tabular}{llccc}
    \hline
    function & concept words                & number & from & focus \\
    \hline
    \sort    & sort                         & 86 & 87 & 0.99 \\
    \prim    & is + prime                   & 83 & 91 & 0.91 \\
    \var     & variance                     & 57 & 67 & 0.85 \\
    \rtrig   & right$|$90$^\circ$ + triangle & 35 & 48 & 0.73 \\
    \arith   & arithmetic + series$|$progression$|$sequence + sum & 33 & 48 & 0.69 \\
    \parln   & parallel                     & 40 & 61 & 0.66 \\
    \hline
    \end{tabular}
\end{table*}

Feitelson et al.\ define two metrics for naming \cite{feitelson:names}:
\emph{focus} is the relative popularity of the most commonly used name, and \emph{diversity} is the tendency to use multiple different names.
A high focus can mean that the situation is clear-cut, and many or most of the experimental subjects understand it the same way --- and consequently use the same name.
But there is always the possibility of small variations.
We therefore define a new metric which we call \emph{semantic focus}.
We first identify the main repeated concepts that appear in the names given to the different functions.
The semantic focus is then the fraction of names that contain all these concepts, implying that semantically they are essentially the same even if syntactically they are not.

The results of performing this analysis on the names given in our experiment are shown in Table \ref{tab:focus}.
The three highest-focus functions are \sort, \prim, and \var.
The semantic focus values for these functions were above 0.85.
The other three functions, \rtrig, \arith, and \parln, had focus in the range 0.66--0.73.

We also note that there are substantial differences in the number of answers given to different questions.
As the order of the questions was randomized, this might imply that subjects found some questions harder than others and skipped them.
In general, the high focus questions were also those that received more answers, which agrees with the conjecture that they were easier.

\section{Discussion}

\subsection{Introducing Intermediate Variables}

Our intention in using version 2 --- introducing intermediate variables but giving them meaningless names --- was to enable a distinction between the effect of the code structure (compound expressions vs.\ multiple simpler expressions) and the naming.
Assuming some version is in general more comprehensible than another, then one might expect that this version would take a shorter time to comprehend for all functions, and that subjects would make fewer mistakes in their understanding.
The assumption implies that the relation between the experimental subjects' performance on the different versions would be consistent: for example, performance on version 2 would be between those of version 1 and version 3.
In particular, we expected that it would be consistently closer to version 1 \emph{or} to version 3.

The results indicate that our assumptions were wrong in this respect.
In terms of correctness, in 2 functions version 2 led to a lower fraction of correct answers than versions 1 and 3 (\arith\ and \parln), while in another it was in between (\rtrig).
In terms of time to correct answer there was one case where the distribution for version 2 was similar to the distribution of version 3 (\rtrig), and several where it was more similar to the distribution of version 1 (\prim, \var, and \parln).

These findings lead to two conclusions.
First, understanding one version is not always easier than understanding another version --- it depends on the function.
Second, the two metrics --- correctness and time to answer correctly --- do not necessarily always correlate with each other.
They appear to measure somewhat different aspects of comprehension performance.

\subsection{Using Meaningful Names}

It is well established that meaningful variable names contribute to program comprehension, and our results corroborate this.
In 5 of 6 cases version 3 of the functions was better understood, sometimes by a substantial margin.
The distribution of time to understand version 3 was either similar to the distributions for other versions, or dominated by them.

Based on these results we can suggest that using intermediate variables \emph{with meaningful names} may be a generally preferred practice.
It either does not matter, or it improved the code's comprehensibility.
However, at present this conclusion rests on a relatively modest experiment.

\subsection{Function Complexity}

The functions we used in the experiment had different levels of complexity.
This does not necessarily refer to some absolute metrics of complexity.
It may also reflect subjective complexity, e.g.\ when some participants find a function harder because they lack some necessary background knowledge.

\begin{table}\centering
    \caption{Metrics reflecting the difficulty of understanding version 1 of different functions, and the improvement when using version 3.}
    \label{tab:func-hard}
    \begin{tabular}{l|cc|cc}
    \hline
    & \multicolumn{2}{c|}{median time [sec]} & \multicolumn{2}{c}{correctness [\%]} \\
    function  & v1  & v3--v1 & v1   & v3--v1 \\
    \hline
    \sort     &  51 &   10 & 94.7\% & --0.8\% \\
    \prim     &  41 &  --6 & 85.7\% &  1.1\% \\
    \var      &  81 & --17 & 64.7\% &  3.0\% \\
    \arith    & 102 &  --7 & 52.8\% &  4.4\% \\
    \parln    & 123 & --27 & 61.1\% & 12.7\% \\
    \rtrig    & 123 & --44 & 31.0\% & 41.8\% \\
    \hline
    \end{tabular}
\end{table}

Table \ref{tab:func-hard} shows two metrics for the difficulty of understanding the different functions, and how it changed between the versions.
The first metric is the median time to understand version 1 of the function.
We use the median rather than the average time because the distributions tend to be skewed, as seen in Fig.\ \ref{fig:cfds}.
To estimate the change between versions 1 and 3 we use the estimated shift in location included in the results of the Mann Whitney procedure used to compare the two distributions.
This is preferable to the difference between the medians since the distributions may display local fluctuations.

The second metric is the percentage of names given to version 1 that were judged to be correct.
The change between version 1 and version 3 is in this case just the difference in percentage points.

As seen in Table \ref{tab:func-hard}, the harder version 1 of a function was --- as reflected by taking longer to comprehend and leading to more mistakes --- the bigger the improvement when intermediate variables with meaningful names were introduced.
The biggest improvement was achieved for the \rtrig\ function.
This is also the only improvement that was found to be statistically significant for both metrics.

At the other end of the scale, the two easiest functions were \prim\ and \sort.
These functions were understood quickly, with few mistakes, and adding intermediate variables did not cause much of a change.
When writing version 3 of \sort\ we chose to define a Boolean intermediate variable named \code{isGreater} which indicated whether the next element in the array is greater than the current one.
In the next line there’s an \code{if} clause consisting only of that variable.
In retrospect we believe that this is a rather artificial intermediate variable, and that most programmers would not use such a variable.
Also, when iterating on the array, we used a \code{rightBoundary} variable to hold the last index of the iteration.
This form of code is less ``natural'' for programmers, especially with python familiarity.
So version 3 was not really better than version 1.

\section{Threats to validity}

We noticed a few threats to the validity of our research which we could mitigate to different degrees.

One of them is a possible survivorship bias.
Participants may quit the experiment at any time or skip a question.
In particular, they may do so if they encounter a difficult task.
This means that the distribution of the answers is not uniform, with the harder questions receiving fewer answers.
Such a bias may lead to inaccurate conclusions.
Note, however, that this bias implies that the results are conservative, as the harder questions are answered preferentially by the stronger participants.
We mitigated this threat to some degree by conducting a pilot study and eliminating the hardest questions, and by reducing the length of the experiment.
It is expected that these actions reduced the dropout rate of participants.

An underlying assumption of most experimental research on comprehension is that longer times and more errors reflect difficulties in comprehension.
While we attempted to use questions that do not require any prior domain knowledge, this is based on an assumption of what people know.
For example, some of the wrong answers about the \rtrig\ function or the \parln\ function might stem from confusion about their geometrical properties rather than from difficulties understanding the code.
To reduce such effects we used the pilot study to filter out candidate questions that were found to be problematic, such as the function calculating the cosine similarity between two vectors.
We also accepted answers that reflect basic understanding of the code even if they did not get all the details right, for example identification of \var\ as calculating the standard deviation, or claims that \parln\ identifies a parallelogram rather than two parallel lines.

Finally, all our functions are basic mathematical functions.
This choice was natural given the research questions, as we need functions that may be written as compound expressions.
However, it may cause an external validity threat, and limit the generalizability of our results.
It is therefore desirable to repeat this line of research using a more diverse set of functions, to check whether similar effects exist in other domains, and also to better characterize the distinction between easy and hard cases.

\section{Conclusions}

Style wars like where to put curly braces are unlikely to be settled.
The problem is that this is mainly a matter of taste and habit, and these factors overshadow any technical differences.
But more involved cases do in fact contain technical substance.
We believe that variable extraction is one such case.
While taste and style may enter into it, there is also a real difference that can be measured experimentally.

We make two main contributions.
The first is to establish an experimental framework for studying this.
This framework is based on selecting specific functions, and creating 3 distinct versions of each one: using a compound expression, partitioned into simpler sub-expressions connected by intermediate variable with meaningless names, and likewise with meaningful names.

The second contribution is our results from applying this framework to a set of basic mathematical functions, and how they were understood by our experimental subjects.
The results suggest that test cases can be roughly divided into two:
simple cases where the intermediate variables do not make much of a difference, and more complicated cases where they do.
In the complicated cases we find that the mere existence of the intermediate variables is not necessarily enough, and may even make things worse.
Improved performance depends on giving them meaningful names.
This can be interpreted as meaning that dividing the problem of comprehending a compound expression into smaller problems does not help so much.
The decisive factor is elevating the level of abstraction by giving the components appropriate names.

An immediate implication of our results is that aggressive mechanized solutions like Spartanization \cite{gil17c} are ill-advised.
The motivation for Spartanization is the desire to find a simple all-encompassing solution to the question of coding style.
Our results imply that there is no fit-all solution, and moreover, that the solution preferred by Spartanization --- to inline all intermediate variables leading to dense compound expressions --- will often be a step in the wrong direction.
For simple cases it does not make much of a difference, but for the complex ones extracting explanatory variables is probably better.
Taken together this implies that if a single rule is sought after, it should be that practitioners should avoid compound expressions and prefer to break them up into simpler sub-expressions with meaningful names.

Note, however, that these results are based on a single modest experiment based on only 6 functions from one domain.
Multiple reproductions, with many different test cases, are required to establish a deeper understanding of the factors that influence the degree to which intermediate variables indeed help to comprehend code.

\section*{Experimental Materials}

Experimental materials are available at\\ {\centering {https://doi.org/10.5281/zenodo.4619665.}

\bibliographystyle{myabbrv}
\bibliography{abbrv,se,refs}

\begin{thebibliography}{10}\itemsep 0pt
\newcommand{\enquote}[1]{``#1''}
\providecommand{\url}[1]{\texttt{#1}}
\providecommand{\urlprefix}{URL }
\expandafter\ifx\csname urlstyle\endcsname\relax
  \providecommand{\doi}[1]{\textsf{\small DOI:\discretionary{}{}{}#1}}\else
  \providecommand{\doi}{\textsf{\small DOI:\discretionary{}{}{}\begingroup
  \urlstyle{rm}\Url}}\fi

\bibitem{ajami19}
S.~Ajami, Y.~Woodbridge, and D.~G. Feitelson, \enquote{\textsl{Syntax,
  predicates, idioms --- what really affects code complexity?}}
  \textit{Empirical Softw.\ Eng.} \textbf{24(1)}, pp.\ 287--328, Feb 2019,
  \doi{10.1007/s10664-018-9628-3}.

\bibitem{arnaoudova14}
V.~Arnaoudova, L.~M. Eshkevari, M.~Di~Penta, R.~Oliveto, G.~Antoniol, and Y.-G.
  Gu\'eh\'eneuc, \enquote{\textsl{{REPENT}: Analyzing the nature of identifier
  renamings}}. \textit{IEEE Trans.\ Softw.\ Eng.} \textbf{40(5)}, pp.\
  502--532, May 2014, \doi{10.1109/TSE.2014.2312942}.

\bibitem{avidan17}
E.~Avidan and D.~G. Feitelson, \enquote{\textsl{Effects of variable names on
  comprehension: An empirical study}}. In 25th \textit{Intl.\ Conf.\ Program
  Comprehension}, pp.\ 55--65, May 2017, \doi{10.1109/ICPC.2017.27}.

\bibitem{binkley15}
D.~Binkley and D.~Lawrie, \enquote{\textsl{The impact of vocabulary
  normalization}}. \textit{Software: Evolution \& Process} \textbf{27(4)}, pp.\
  255--273, Apr 2015, \doi{10.1002/smr.1710}.

\bibitem{binkley09c}
D.~Binkley, D.~Lawrie, S.~Maex, and C.~Morrell, \enquote{\textsl{Identifier
  length and limited programmer memory}}. \textit{Sci.\ Comput.\ Programming}
  \textbf{74(7)}, pp.\ 430--445, May 2009, \doi{10.1016/j.scico.2009.02.006}.

\bibitem{campbell07}
I.~Campbell, \enquote{\textsl{Chi-squared and {Fisher–Irwin} tests of
  two-by-two tables with small sample recommendations}}. \textit{Statist.\
  Med.} \textbf{26}, p.\ 3661–3675, 2007, \doi{10.1002/sim.2832}.

\bibitem{caprile00}
B.~Caprile and P.~Tonella, \enquote{\textsl{Restructuring program identifier
  names}}. In \textit{Intl.\ Conf.\ Softw.\ Maintenance}, pp.\ 97--107, Oct
  2000, \doi{10.1109/ICSM.2000.883022}.

\bibitem{counsell15}
S.~Counsell, X.~Liu, S.~Swift, J.~Buckley, M.~English, S.~Herold, S.~Eldh, and
  A.~Ermedahl, \enquote{\textsl{An exploration of the `introduce explaining
  variable' refactoring}}. In \textit{XR '15 Workshops}, art.\ no.~9, May 2015,
  \doi{10.1145/2764979.2764988}. (Intl.\ Workshop Refactoring \& Testing).

\bibitem{deissenboeck05}
F.~{Dei{\ss}enb\"ock} and M.~Pizka, \enquote{\textsl{Concise and consistent
  naming}}. In 13th \textit{IEEE Intl.\ Workshop Program Comprehension}, pp.\
  97--106, May 2005, \doi{10.1109/WPC.2005.14}.

\bibitem{fakhoury18}
S.~Fakhoury, Y.~Ma, V.~Arnaoudova, and O.~Adesope, \enquote{\textsl{The effect
  of poor source code lexicon and readability on developers' cognitive load}}.
  In 26th \textit{Intl.\ Conf.\ Program Comprehension}, pp.\ 286--296, May
  2018, \doi{10.1145/3196321.3196347}.

\bibitem{feitelson:names}
D.~G. Feitelson, A.~Mizrahi, N.~Noy, A.~Ben~Shabat, O.~Eliyahu, and R.~Sheffer,
  \enquote{\textsl{How developers choose names}}. \textit{IEEE Trans.\ Softw.\
  Eng.} \doi{10.1109/TSE.2020.2976920}. (early access).

\bibitem{fowler:refactor}
M.~Fowler, \textit{Refactoring: Improving the Design of Existing Code}. Pearson
  Education, Inc., 2nd ed., 2019.

\bibitem{gil17b}
Y.~Gil and M.~Orr\`u, \enquote{\textsl{The {Spartanizer}: Massive automatic
  refactoring}}. In 24th \textit{IEEE Intl.\ Conf.\ Softw.\ analysis,
  Evolution, \& Reengineering}, pp.\ 477--481, Feb 2017,
  \doi{10.1109/SANER.2017.7884657}.

\bibitem{gil17c}
Y.~Gil and M.~Orr\`u, \enquote{\textsl{Code {Spartanization}: One rational
  approach for resolving religious style wars}}. In \textit{Symp.\ Applied
  Computing}, pp.\ 1615--1622, Apr 2017, \doi{10.1145/3019612.3019748}.

\bibitem{hofmeister19}
J.~C. Hofmeister, J.~Siegmund, and D.~V. Holt, \enquote{\textsl{Shorter
  identifier names take longer to comprehend}}. \textit{Empirical Softw.\ Eng.}
  \textbf{24(1)}, pp.\ 417--443, Feb 2019, \doi{10.1007/s10664-018-9621-x}.

\bibitem{holzmann16}
G.~J. Holzmann, \enquote{\textsl{Code clarity}}. \textit{IEEE Softw.}
  \textbf{33(2)}, pp.\ 22--25, Mar/Apr 2016, \doi{10.1109/MS.2016.44}.

\bibitem{lawrie06}
D.~Lawrie, C.~Morrell, H.~Field, and D.~Binkley, \enquote{\textsl{What's in a
  name? a study of identifiers}}. In 14th \textit{Intl.\ Conf.\ Program
  Comprehension}, pp.\ 3--12, Jun 2006, \doi{10.1109/ICPC.2006.51}.

\bibitem{rajlich97}
V.~Rajlich and G.~S. Cowan, \enquote{\textsl{Towards standard for experiments
  in program comprehension}}. In 5th \textit{IEEE Intl.\ Workshop Program
  Comprehension}, pp.\ 160--161, Mar 1997, \doi{10.1109/WPC.1997.601284}.

\bibitem{scanniello13}
G.~Scanniello and M.~Risi, \enquote{\textsl{Dealing with faults in source code:
  Abbreviated vs.\ full-word names}}. In 29th \textit{Intl.\ Conf.\ Softw.\
  Maintenance}, pp.\ 190--199, Sep 2013, \doi{10.1109/ICSM.2013.30}.

\bibitem{schankin18}
A.~Schankin, A.~Berger, D.~V. Holt, J.~C. Hofmeister, T.~Riedel, and M.~Beigl,
  \enquote{\textsl{Descriptive compound identifier names improve source code
  comprehension}}. In 26th \textit{Intl.\ Conf.\ Program Comprehension}, pp.\
  31--40, May 2018, \doi{10.1145/3196321.3196332}.

\bibitem{wiese19}
E.~S. Wiese, A.~N. Rafferty, and A.~Fox, \enquote{\textsl{Linking code
  readability, structure, and comprehension among novices: It's complicated}}.
  In 41st \textit{Intl.\ Conf.\ Softw.\ Eng.}, pp.\ 84--94, May 2019,
  \doi{10.1109/ICSE-SEET.2019.00017}. (SEET track).

\bibitem{zabardast20}
E.~Zabardast, J.~Gonzalez-Huerta, and D.~\v{S}mite,
  \enquote{\textsl{Refactoring, bug fixing, and new development effect on
  technical debt: An industrial case study}}. In 46th \textit{Euromicro Conf.\
  Softw.\ eng.\ \& Advanced Apps.}, pp.\ 376--384, Aug 2020,
  \doi{10.1109/SEAA51224.2020.00068}.

\end{thebibliography}

\end{document}